# Nanocommunication via FRET with DyLight Dyes using Multiple Donors and Acceptors

Kamil Solarczyk, Krzysztof Wojcik, and Pawel Kulakowski

*Abstract*—The phenomenon of Förster Resonance Energy Transfer, commonly used to measure the distances between fluorophore molecules and to study interactions between fluorescent-tagged proteins in life sciences, can also be applied in nanocommunication networks to transfer information bits. The mechanism offers a relatively large throughput and very small delays, but at the same time the channel bit error rate is too high and the transmission ranges are too limited for communication purposes. In this paper, multiple donors at the transmitter side and multiple acceptors at the receiver side are considered to decrease the bit error rate. As nanoantennas, the DyLight fluorescent dyes, which are very well suited to long range nanocommunication due to their large Förster distances and high degrees of labeling, are proposed. The reported results of the recent laboratory experiments confirm efficient communication on distances over 10 nm.

## I. INTRODUCTION

THE current development of nanomachines and nanorobots and organizing them into systems is stimulated by incredible applications promised by them in industrial manufacturing, biology and, especially, medicine. The progress in the latter field may be revolutionized with swarms of tiny robots employed in targeted drug delivery, *in vivo* imaging and diagnostics, tissues regeneration and engineering [1, 2]. The future success of nanomachine systems depends, however, on overcoming two challenges: the first one is fabrication of nanodevices with proper precision, lifetime and efficiency [3]. The second challenge, being the subject of this paper, is the efficient communication between the nanorobots: they obviously need to signal their actions and communicate with each other.

The manuscript was submitted September 30, 2015. The work was performed under Contract 11.11.230.018 and also funded by the National Science Centre based on the decision number DEC-2013/11/N/NZ6/02003. The confocal microscope was purchased through an EU structural funds grant BMZ no. POIG.02.01.00-12-064/08.

K. Wojcik and K. Solarczyk are with Division of Cell Biophysics, Faculty of Biochemistry, Biophysics and Biotechnology, Jagiellonian University, 7, Gronostajowa St., 30-387 Kraków, Poland (e-mails: krzysztof.wojcik@uj.edu.pl, kj.solarczyk@uj.edu.pl). K. Wojcik is also with 2nd Department of Medicine, Jagiellonian University Medical College, Kraków, Poland.

P. Kulakowski is with Department of Telecommunications, AGH University of Science and Technology, Al. Mickiewicza 30, 30-059 Kraków, Poland (e-mail: kulakowski@kt.agh.edu.pl).

The common mechanisms considered in nanocommunication literature are: calcium ion signaling [4-5], flagellated bacteria carrying data in its DNA [6], the movement of molecular motors [7-8], pheromones, pollen and spores [9]. They are, however, based on mechanical phenomena and thus their propagation delays and achievable data throughputs are not satisfactory for communication purposes. The delay issues were the reason to propose acoustic communication techniques for nanoscale [10]. Here, we consider even more rapid phenomena, namely Förster Resonance Energy Transfer (FRET).

FRET has been already proposed as an efficient means for nanoscale communication [9,11-12]. It is a non-radiative process of transferring energy from an excited fluorophore molecule called the donor to an adjacent molecule called the acceptor, being in the ground state. Popular molecules that can be donors and acceptors in a FRET process are: fluorescent dyes, proteins and quantum dots. Fluorescent dyes, which are quite small structures of about few nanometers, can serve as nanoantennas being attached to larger nanomachines and performing numerous tasks, e.g. transporting cargo (kynesins, dyneins) or seeking other molecules (antibodies), see Fig. 1. FRET is characterized by a very small delay, about 10-20 nanoseconds, and potentially a very high throughput of tens of Mbit/s. The main drawbacks of the FRET mechanism are a very high channel bit error rate and an effective transmission range limited to a few nanometers. It has been also already shown that the channel bit error rate can be decreased when multiple donors at the transmitter side and multiple acceptors at the receiver side are applied to create so called MIMO (multiple-input multiple-output) FRET channel [13].

The limited transmission range, however, remains an Achilles heel of the FRET-based communication technique. The FRET efficiency decreases with sixth power of the donor-acceptor separation (see Eq. 1 in the following section), what means that increasing the transmission distance is much harder than in wireless communication. In this paper, we report experiments on DyLight dyes, which is quite a new family of fluorophores much better suited for future nanocommunication applications. We also consider multiple donors and acceptors here, constructing 4 different nanonetworks based on DyLight dyes. On the basis of a laboratory measurements campaign, we report successful communication over 12-13 nm, which is a 50% increase of transmission range comparing with previous experiments. We expect it will be crucial for the efficient cooperation of future nanomachines. Many nanomachines, e.g. some antibodies, kynesins, dyneins, are of size close to 10 nm (see Fig. 1), so it would be quite hard for them to approach each other and communicate on distances of only few nanometers.

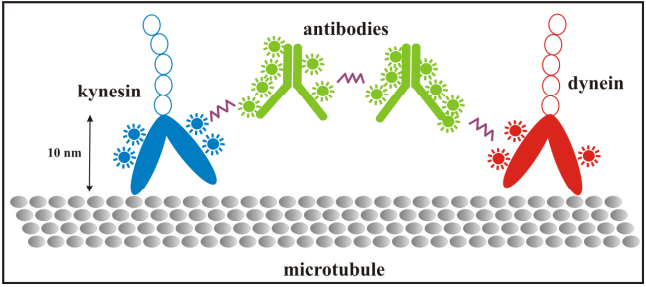

Fig. 1. Communication is nanoscale between different protein-based nanomachines: kynesins and dyneins walking over a microtubule and antibodies. All proteins are labeled with fluorescent dyes (marked as small circles with rays). The FRET process (marked as a violet zigzag) may happen between the dyes on different proteins.

In particular, the contribution of this paper can be summarized as follows:

1. We propose DyLight dyes as efficient candidates for nanoantennas in MIMO-FRET based nanocommunication.

2. We construct four different nanonetworks based on DyLight dyes and perform the first nanocommunication experiments with this fluorescent family.

3. We obtain results showing a 50% increase of communication range comparing with previously reported experiments [13].

4. Additionally, we calculate and present bit error rate curves for the chosen MIMO-FRET channels based on DyLight dyes.

The rest of the paper contains the following sections. In Section II, the principles of FRET theory are introduced together with its extension to the case of multiple donors and multiple acceptors. In Section III, a description is given of nanonetworks built on antibodies and DyLight dyes. Section IV contains the laboratory methodology. The results of the experiments and the calculations are given in Section V. Finally, in Section VI, conclusions and future directions of study on the topic of FRET-based nanonetworks are outlined.

## II. FRET WITH MULTIPLE DONORS AND ACCEPTORS

If we look at the FRET process from a communications point of view we can think of donors and acceptors as transmit and receive antennas, respectively. Consequently, the FRET pair can be considered a wireless communication system. The probability of the energy transfer is described as FRET efficiency and is given by [14]:

$$E = \frac{R_0^6}{r^6 + R_0^6} . \qquad (1)$$

In (1), $r$ is the donor-acceptor separation, while $R_0$, called Förster distance, is the donor-acceptor separation where $E$=50%. This separation is specific for each donor-acceptor pair and can be measured experimentally (usually 3-8 nm). In general, the Förster distance depends on the overlap between the donor emission and acceptor absorption spectra as well as on the quantum yield of the donor. The larger the $R_0$, the better the match of donor and acceptor spectra and the more efficient the FRET process. Förster distances are usually larger when the dyes' spectra are located in higher wavelengths, but this is not a rule, as it depends on the spectra shapes [14]. If the excitation energy is not transferred via FRET, it can be also emitted as a photon or dissipated. These two last phenomena should be interpreted as channel losses increasing the error rate.

The FRET phenomenon is well known in life sciences and is commonly used to measure distances between fluorophores (organic molecules whose emission spectra cover the range of visible light) or to study interactions between fluorescent-tagged proteins. However, it can also be applied to nanocommunications. Let's imagine a molecular structure performing the function of a molecular transmitter. Let's have a donor molecule attached to it, as a molecular transmit antenna. On the other side, there could be another structure: a molecular receiver with an acceptor attached (a molecular receive antenna). Now, information can be sent from the Tx to the Rx side via the FRET channel. This can be simply achieved by exciting the donor when bit '1' is to be sent and by keeping the donor in the ground state when transmitting bit '0'. The FRET delay is relatively small, in the order of nanoseconds. If FRET was efficient in 100% of cases, the associated data throughput (calculated as the inverse of the delay) could then be as high as tens or even hundreds of Mbit/s. However, FRET efficiency is only 50% for the donor and acceptor separated by $R_0$ and it decreases with the sixth power of that distance. This results in quite a high channel bit error rate (BER): transmitting '0' is always successful, but transmitting '1' is erroneous with the probability of 1-$E$. Thus the FRET channel BER is equal to:

$$BER = 0.5 \cdot (1 - E) = \frac{0.5 \cdot r^6}{r^6 + R_0^6} . \qquad (2)$$

Consequently, in order to obtain BER=0.1%, which is a common value in telecommunication standards, the donor-acceptor separation should not exceed $0.355R_0$. This result limits the efficiency of FRET-based nanocommunication to distances below 3.5 nm.

FRET efficiency can be increased, if there is not one but more acceptor molecules in the vicinity of the donor [15]. When an excited donor is surrounded by $m$ equally distant acceptors, the probability that the excitation energy is transferred via FRET to one of the acceptors is given by [14]:

$$E_{1,m} = \frac{m \cdot R_0^6}{r^6 + m \cdot R_0^6} . \qquad (3)$$

This means that having a molecular receiver with multiple acceptor molecules (Rx antennas) can greatly increase FRET efficiency and decrease the channel BER, which is then equal to:

$$BER_{1,m} = \frac{0.5 \cdot r^6}{r^6 + m \cdot R_0^6} . \qquad (4)$$

Furthermore, we can also use molecular transmitters with multiple donors. If there are $n$ donor molecules, we can excite them all when bit '1' is to be transmitted. Then, (again, assuming they are equidistant from the acceptors) the probability that *at least* one of them transfers its excitation energy to an acceptor is much higher. Let us assume the FRET events for all the donors are independent of each other, which is reasonable, as the FRET efficiency does not depend on the fact if there are other excited molecules nearby [14]. Now, the probability that *none* of the donors transfers its energy to any of the acceptors can be expressed as the *n*-th power of such a probability for a single donor:

$$\left(1 - \frac{m \cdot R_0^{\;6}}{r^6 + m \cdot R_0^{\;6}}\right)^n \qquad (5)$$

Thus, the probability that *at least* one donor succeeds in transferring its excitation energy to an acceptor is equal to:

$$E_{n,m} = 1 - \left(1 - \frac{m \cdot R_0^{\;6}}{r^6 + m \cdot R_0^{\;6}}\right)^n. \qquad (6)$$

Consequently, the bit error rate of such a FRET communication channel is much lower than in the basic case (with a single donor and a single acceptor) and can be calculated as:

$$\mathrm{BER}_{n,m} = 0.5 \cdot \left(\frac{r^6}{r^6 + m \cdot R_0^{\;6}}\right)^n. \qquad (7)$$

The number of donors/acceptors attached to a molecular structure is a parameter called the *degree of labeling* (DoL) and it is a parameter not easy to control, as will be further discussed in the following sections. In the area of communications, this reminds the known idea of MIMO (multiple-input multiple-output) channels [16, 17]. Therefore, we will henceforth use the term: MIMO-FRET channels.

## III. BUILDING NANONETWORKS WITH DYLIGHT DYES

The absorption and emission spectra of the DyLight fluorescent dye family series cover much of the visible light and extend into the infrared region (the spectra maxima range from 350 nm to 794 nm), allowing the dyes detection using most types of fluorescence and confocal microscopes, as well as infrared imaging systems. In comparison with other fluorescent dye families (Alexa Fluor, CyDye, LI-COR), the main advantages of the DyLight group are as follows [18, 19]: (a) absorption and emission spectra of a similar shape (resulting in large Förster distances), (b) high photostability and brightness, (c) low pH-sensitivity, (d) high dye-to-protein (degree of labeling) ratio in water solutions to be achieved without precipitation of conjugates. The most relevant measurements on FRET between DyLight dyes reported in the open literature are, until now, focused on describing spectral properties of DyLight fluorophores [19] and using FRET, also with multiple acceptors, in disease diagnostics [20].

For the purpose of studying the signal transfer via FRET between DyLight dyes, we built special molecular structures based on antibodies, i.e. proteins with the ability to recognize and bind other molecules (Fig. 2). When thinking about future applications in nanocommunication, antibodies can be treated as nanomachines and DyLight dyes as nanoantennas attached to them. They may diffuse in a cellular membrane and perform some tasks, e.g. looking for antigens. When two antibodies are close to each other, the communication process may occur. In our case, in order to perform FRET measurements the chosen antibodies were bound to a molecular skeleton consisting of DNA and histone proteins in the nuclei of fixed HeLa cells (Fig. 2). It should be emphasized that our choice of building the network of antibodies in cells was solely technical, i.e.

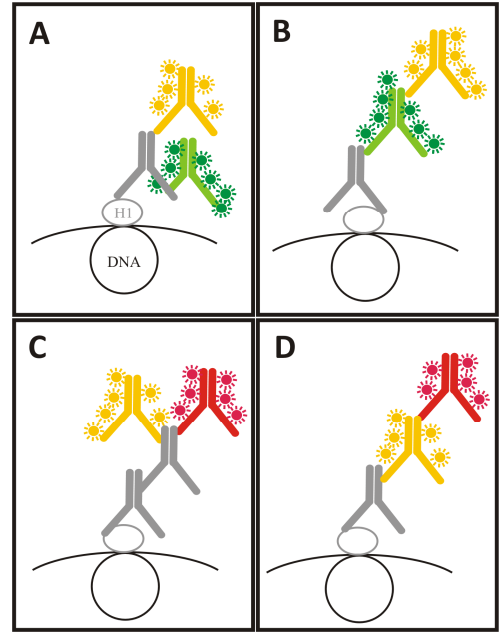

Fig. 2. Four investigated scenarios: molecular structures prepared for the purpose of the experiments. Each structure consists of a DNA molecule, a histone, one or two linker (gray color here) antibodies and finally two more antibodies with donor/acceptor dyes attached. The antibodies labeled with DyLight dyes are marked as green/yellow/red for 549/594/649 dyes respectively.

driven by fact that the nuclei contain a well-known and easily accessible scaffold for binding of antibodies. We investigated 4 different scenarios: for each of them, we had a molecular structure built of antibodies. In each scenario, two of the antibodies were labeled with multiple DyLight dyes: donors on one side and acceptors on the other side, creating a MIMO-FRET communication channel. In order to obtain good communication efficiency and low BER for long transmission ranges the FRET dyes pairs should fulfill the following criteria: (a) the donor emission spectra and acceptor absorption spectra should match each other well, resulting in high values of Förster distances and (b) the degree of labeling of the dyes used should be as high as possible. Thus, we finally chose two pairs:

- DyLight 549 dyes as donors and DyLight 594 as acceptors, donor-acceptor $R_0 = 5.62$ nm,
- DyLight 594 as donors and DyLight 649 as acceptors, donor-acceptor $R_0 = 8.16$ nm.

All the dyes were supplied by Jackson ImmunoResearch Laboratories, Inc. Their numbers, 549, 594 and 649, indicate the main wavelengths of their absorption spectra given in nanometers. Each pair was additionally tested in two spatial configurations, with slightly different donor-acceptor distances and connections of antibodies, resulting in 4 scenarios in total, see Fig. 2. It allowed us to observe the distance-dependent effects, the influence of the choice of antibodies and the variability of possible lifetimes (especially in the scenario C). All the experiments were performed on fixed and permeabilized cells, in order to allow the antibodies to easily enter the nucleus and guarantee that all chemical reactions in the cells were stopped and did not affect the measurements.

The details of the construction of the molecular structures were as follows. The first element of the structure was always a histone H1 bound to a DNA molecule. Next, one or two linker antibodies were attached to the histone H1. Then, an antibody labeled with donor DyLight dyes was bound to the linker ones. Finally, the last antibody labeled with acceptor dyes was attached directly to the one labeled with donors (scenario B and D) or to the linker antibodies (scenarios A and C), as shown in Fig. 2. Each antibody was a molecule of Immunoglobulin G, being like a 3-element airscrew in shape [21] with a radius of about 7 nm.

TABLE I
SEPARATIONS BETWEEN THE ANTIBODIES, DONOR AND ACCEPTOR DYES

| scenario | A | B | C | D |
|---|---|---|---|---|
| antibody on Tx side | Goat Anti-Mouse IgG 115-505-008 | Goat Anti-Mouse IgG 115-505-008 | Rabbit Anti-Goat IgG, Fc Fragment Specific 305-515-046 | Goat Anti-Mouse IgG 115-515-062 |
| antibody on Rx side | Goat Anti-Mouse IgG 115-515-062 | Rabbit Anti-Goat IgG, Fc Fragment Specific 305-515-046 | Rabbit Anti-Goat IgG, Fc Fragment Specific 305-495-008 | Rabbit Anti-Goat IgG, Fc Fragment Specific 305-495-008 |
| dyes on Tx side (donors) | DyLight 549 | DyLight 549 | DyLight 594 | DyLight 594 |
| dyes on Rx side (acceptors) | DyLight 594 | DyLight 594 | DyLight 649 | DyLight 649 |
| separation between antibodies (centre to centre) | 8-10 nm | 11-13 nm | 12-14 nm | 11-13 nm |

In the labeling process, it is not possible to choose exact positions of the dyes on the antibodies; the donors and acceptors (DyLights 549, 594 and 649) can be attached to the respective antibodies on their whole length[1]. Still, we have constructed the antibodies chains in a way to have relatively large distances between the donors and acceptors (on average 9 nm in scenario A and 12-13 nm in scenarios B-D). On the basis of the known antibody geometries [21], we can assess the separations between the antibodies where the dyes were attached (see Table I). These separations are the average ones.

Due to the distribution of the dyes on the surface of antibodies, the real donor-acceptor distances may vary from these estimated values by a few nanometers. Moreover, we cannot exclude a slight rotation or bending of the antibodies, what can change the final donor-acceptor distances by an additional 1 nm. As the FRET probability decreases with the sixth power of this distance, we expect that FRET happens mainly between the donors and acceptors which are closest to each other in the moment of the excitation of the donor. It is another reason why the high DoL values are so important for effective FRET-based nanocommunication.

According to the information provided by Jackson ImmunoResearch Laboratories, the average degrees of labeling for the dyes are:
• DyLight 549: DoL = 7
• DyLight 594 in scenarios A and C: DoL = 6.4
• DyLight 594 in scenarios B and D: DoL = 5.2
• DyLight 649: DoL = 5.5[2]

These DoL values are not always integer numbers, which means there are two possibilities of how many dyes are attached to an antibody (e.g. 5 or 6, 6 or 7) and the resulting DoL is the weighted mean of two numbers.

## IV. LABORATORY METHODOLOGY

### A. Cell culture

In order to allow their growth, HeLa cells were maintained in Dulbecco's Modified Eagle Medium (containing all necessary nutrients) supplemented with 10% fetal bovine serum at 37°C and 5% $CO_2$. 24-48h before the experiment, the cells were seeded on 18mm microscope slides. Before immunolabeling, the cells were fixed in 4% formaldehyde, permeabilized in 0.1% Triton X-100 and then blocked in 3% bovine serum albumin (BSA). The fixed cells were incubated with the primary antibody for 1h in 3% BSA at room temperature, washed three times in phosphate-buffered saline (PBS) and then incubated overnight at 4°C with an appropriate secondary (donor dye labeled) antibody. After performing lifetime measurements of the structures with antibodies labeled with the donor dye only, incubation with another antibody (acceptor dye labeled) was performed at the microscope stage in order to establish the required geometrical configuration of the dyes.

### B. FLIM measurements

FRET efficiency values cannot be measured directly. Instead, two techniques are mainly used. The first one takes advantage of the fact that fluorescence intensity of the donor should be decreased in the presence of an acceptor. However, accurate fluorescence intensity measurements are often impeded by spectral bleed-through and photobleaching of the dyes. The second technique is based on the measurements of fluorescence lifetimes of fluorophores. Fluorescence lifetime, described as the average time a fluorophore spends in the excited state before emitting a photon, is less influenced by photobleaching and thus allows to measure FRET more accurately. Because the lifetime of a donor should be shortened in the presence of an acceptor, two measurements are required: one before and one after the addition of the acceptor. Then, the FRET efficiency can be calculated according to the following formula [14]:

$$E = 1 - \frac{\tau_{with-acceptors}}{\tau_{no-acceptors}}, \qquad (8)$$

where $\tau_{no-acceptors}$ and $\tau_{with-acceptors}$ are the measured lifetimes of the donor in the absence and presence of an acceptor, respectively. In our experiments, to measure fluorescence lifetimes of the DyLight dyes we used a Leica TCS SP5 II SMD confocal microscope (Leica Microsystems GmbH) integrated with FCS/FLIM TCSPC module from PicoQuant GmbH (Fig. 3). The FLIM (Fluorescence Lifetime Imaging Microscopy) module integrated with a confocal microscope permits to record microscope images that contain information about fluorescence intensity and lifetime in every pixel of the image. In all measurements, the FLIM image was captured at 256x256 pixel resolution and at a speed of 200 lines/s. The acquisition time for each image was set to 1 minute with a laser repetition rate of 40 MHz. The laser power was set to achieve a photon counting rate of 200-300 kCounts/s. In order to measure FRET efficiency between DyLight dyes, two lifetime images were collected for each scenario, one before and one after the incubation with an acceptor-labeled antibody. Both DyLight 549 (donors in scenarios A and B) and DyLight 594 (donors in scenarios C and D) were excited with 470 nm laser line and the emission was collected with 500-560 and 607-683 nm band-pass filters, respectively. Three images of approximately 50x50 um were collected for each pair, with 4-6 cells visible in this region. For the purposes of the analysis, only the nuclei of the cells were selected in order to exclude any signal resulting from non-specific binding of the antibodies. The collected data was analyzed using SymPhoTime II software (PicoQuant GmbH), excluding the expected instrument response function time range (tail-fitting method). The DyLight 549 and DyLight 594 lifetime decays were fitted with a two-exponential function:

$$F(t) = A_1 \cdot e^{-t/\tau_1} + A_2 \cdot e^{-t/\tau_2}, \qquad (9)$$

where $\tau_1$ and $\tau_2$ are the lifetimes, while $A_1$ and $A_2$ represent the amplitudes of the components. The final lifetime was calculated as the amplitude-weighted average of $\tau_1$ and $\tau_2$:

$$\tau = \frac{\sum_i A_i \cdot \tau_i}{\sum_i A_i}. \qquad (10)$$

---

[1] The positions of the dyes could be determined more precisely if smaller molecules were used instead of Immunoglobulin G (IgG). For instance, the donor and acceptor dyes could also be attached to fragment antigen-binding (Fab) molecules, which are 3 times smaller than IgG [22]. The dyes on a Fab particle would be located closer to each other, but having in mind smaller Fab size comparing with IgG, it would be more difficult to obtain a high degree of labeling.

[2] These DoL are quite high comparing with typical values reported in fluorescence spectroscopy experiments. We have chosen these fluorophores also because of their high possible DoL, as this factor determines the FRET efficiency. The DyLight dyes with lower DoL could also be used here, but, according to (5) and (6), the communication channel would have poorer characteristics.

The goodness-of-fit was estimated based on the weighted residuals and the chi squared value.

### C. Image acquisition

Steady-state (containing information about fluorescence intensity only) images of the cells were obtained using a Leica TCS SP5 II SMD confocal microscope equipped with a 63x NA 1.4 oil immersion lens. All images were captured at a 512x512 pixel resolution, with a scanning speed of 200 lines/s and with 2 frames averaged. DyLight 549 was excited with a 488 nm laser line from an argon laser, while its emission was collected in a photo-multiplier tube (PMT) at 560-580 nm. DyLight 594 was excited with a 594 laser line from a HeNe laser and its emission collected at 600-620 nm. In analogy to FLIM measurements, two images were collected for each spatial configuration: one before and one after incubation with the donor-labeled antibody. The equal laser power and detector gain were used for each pair of the dyes.

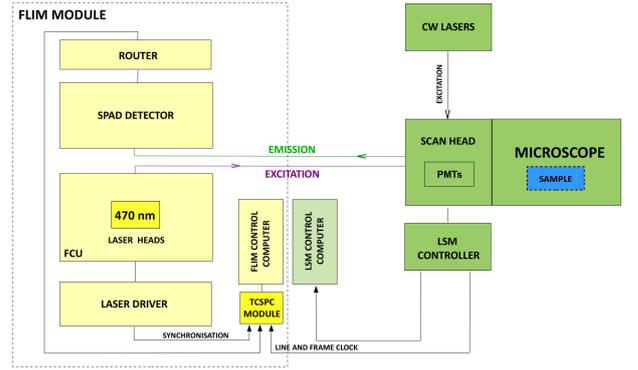

Fig. 3. Schematic representation of the Laser Scanning Microscope integrated with a FLIM module. The experiment begins when the sample (HeLa cells) is excited with a pulsed laser (470 nm) controlled by a laser driver. This unit allows us to control the laser power output and its repetition rate. Upon excitation, the sample emits fluorescence photons which are detected by SPAD detectors. In order to determine the exact time between the excitation pulse and the arrival of the first photon at the detector, a TCSPC module connected to a computer receives information from the detectors, laser driver and laser scanner. In steady-state measurements (image acquisition without information about fluorescence lifetimes), the sample is excited with a continuous wave laser, the signal is detected in PMT detectors and the resulting image is viewed on a LSM computer.

## V. EXPERIMENT RESULTS

Finally, for each of the scenarios A, B, C and D, we performed the FLIM measurements for 13-17 nuclei of HeLa cells. Every nucleus was analyzed separately. We estimated that there were at least $3 \cdot 10^6$ constructed molecular structures (a DNA molecule with a histone H1, antibodies labeled with donor and acceptor dyes) with MIMO-FRET channels in every nucleus. For each nucleus, we measured the fluorescence lifetimes of the acceptor molecules without and in the presence of the donors (see Table II). The sets of data together with fitting curves for all 4 scenarios are plotted in Figure 4. These are the examples for the chosen nuclei. The lifetimes may vary from one nucleus to another, as they are influenced by the environment of the fluorophore (the density of molecules not participating in the FRET process, like other proteins and DNA) [23]. The lifetimes for the cases without and with the donors are, however, usually proportional, what means that the FRET efficiency, calculated with equation (8), is stable. The average FRET efficiencies are given in Table III.

Knowing the laser repetition rate (40 MHz) and the achieved photon counting rate (200-300 kCounts/s), and bearing in mind that the measured FRET efficiencies did not exceed 50% (see the results below), we clearly see that we had only isolated donor dyes excited (or no excitation at all) per laser pulse. As each HeLa nucleus contained millions of donor dyes, it was not possible to excite all the donor dyes simultaneously and validate the full MIMO-FRET transmission scheme. Instead, we measure the MIMO (1,*m*) case, i.e. with a single donor excited and multiple acceptors being able to receive the FRET energy.

| Scenario | | | | | | | |
|---|---|---|---|---|---|---|---|
| A | | B | | C | | D | |
| no/with acceptors | | no/with acceptors | | no/with acceptors | | no/with acceptors | |
| 0.81 | 0.59 | 0.80 | 0.76 | 2.10 | 1.51 | 2.03 | 1.35 |
| 0.8 | 0.58 | 0.77 | 0.75 | 2.12 | 1.48 | 2.00 | 1.37 |
| 0.82 | 0.59 | 0.84 | 0.80 | 2.11 | 1.52 | 2.04 | 1.41 |
| 0.83 | 0.59 | 0.80 | 0.77 | 2.15 | 1.50 | 2.07 | 1.43 |
| 0.75 | 0.6 | 0.78 | 0.75 | 2.12 | 1.57 | 2.03 | 1.39 |
| 0.82 | 0.62 | 0.80 | 0.72 | 2.17 | 1.51 | 2.20 | 1.25 |
| 0.83 | 0.61 | 0.81 | 0.74 | 2.32 | 1.51 | 2.22 | 1.27 |
| 0.84 | 0.62 | 0.83 | 0.74 | 2.25 | 1.49 | 2.22 | 1.24 |
| 0.79 | 0.6 | 0.83 | 0.75 | 2.77 | 1.65 | 2.19 | 1.21 |
| 0.8 | 0.61 | 0.75 | 0.70 | 2.79 | 1.75 | 2.18 | 1.26 |
| 0.75 | 0.59 | 0.75 | 0.68 | 1.95 | 1.15 | 2.21 | 1.27 |
| 0.8 | 0.61 | 0.75 | 0.71 | 1.87 | 1.13 | 2.22 | 1.26 |
| 0.73 | 0.58 | 0.79 | 0.73 | 2.07 | 1.24 | 2.20 | 1.23 |
| 0.78 | 0.6 | 0.73 | 0.68 | 1.83 | 1.09 | | |
| | | 0.77 | 0.71 | 2.05 | 1.24 | | |
| | | | | 2.06 | 1.14 | | |
| | | | | 1.95 | 1.17 | | |
| Average lifetimes and their standard deviations for each scenario | | | | | | | |
| 0.8 ± 0.03 | 0.6 ± 0.01 | 0.79 ± 0.03 | 0.73 ± 0.03 | 2.16 ± 0.26 | 1.39 ± 0.21 | 2.14 ± 0.09 | 1.3 ± 0.08 |

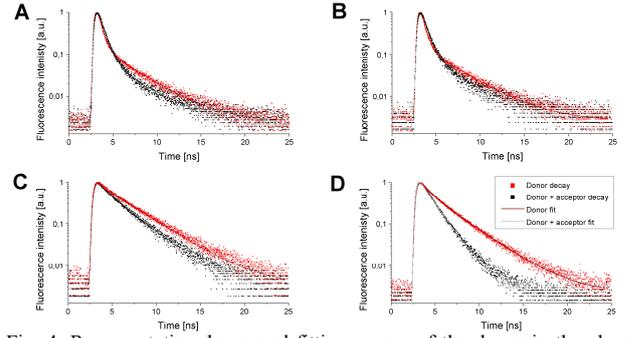

Fig. 4. Representative decay and fitting curves of the donor in the absence and presence of the acceptor measured in HeLa nuclei for scenarios A, B, C and D.

The measured FRET efficiencies together with confidence intervals for all 4 scenarios are shown in Table III. For each FRET efficiency, we also give the respective BER values, calculated as: $0.5 \cdot (1 - E)$. At the same time, the FRET phenomenon was observed in microscope images as a decrease in donor fluorescence intensity after adding the acceptor dyes, see Fig 5.

The measurement results can be verified by FRET theory. Assuming the average donor-acceptor separation equal to the distance between the respective antibodies (see Table I) and putting it together with the known $R_0$ into the equation (3), we can calculate theoretical values of $E_{(1,m)}$. The calculations must be performed twice, as we do not know the exact number of acceptors: in scenario A we had 6 or 7 acceptors (DoL = 6.4), in scenarios B-D we had 5 or 6 acceptors (DoL = 5.2 and 5.5). Hence, we obtain a theoretical range for $E_{(1,m)}$: it is given in the last row of Table III.

The measurement results match the theoretical ones quite well, except in scenario C. In all scenarios, the real donor-acceptor separations may vary, as we have no control regarding exact positions of the DyLight dyes. Furthermore, the antibodies chain shape depends on the axial (0-180 deg) and segmental flexibility (0-180 deg); additionally the switch peptide (elbow) also increases the flexibility of the antibody Fab region (0-50 deg) [24], which may influence the donor-acceptor separation. In scenario C, we assumed the labeled antibodies bound the linker antibody in the best way from the energetic point of view, which meant the distance between the labeled antibodies was maximal. However, in practice, they may bind closer to each other; thus the donor-acceptor separation is smaller than predicted in Table III. As the FRET efficiency is inversely proportional to the sixth power of this separation, these effects cause higher than expected FRET efficiency.

The experiments were performed for relatively long distances between the Tx and Rx sides, i.e. with the donor-acceptor separations significantly exceeding their Förster distances. Thus, as we can see from Table III, even with a large number of acceptor molecules, it was not possible to obtain FRET efficiency higher than 50%. Also the respective bit error rates (30-50%) are highly non satisfactory for telecommunication purposes. A solution for this (other than exciting the donor multiple times per bit, what however decreases the corresponding data throughput [25]) might lie in exploiting full MIMO communication, i.e. to excite all the donor dyes simultaneously. It might be realized with a very strong external laser impulse or using the energy of a local chemical reaction, in this latter case it is called Bioluminescence Resonance Energy Transfer. The receiver decodes the bit as '1' if any of acceptors is excited, i.e. if at least one donor transfers its excitation energy to at least one acceptor. If no acceptors are excited, the bit is decoded as '0'. For the proper data transmission, it is crucial to correctly choose the rate (frequency) of sending bits. Obviously, from the telecommunication point of view the higher rate the better, but the FRET and fluorescence delays are going to put the limits here. As documented in Table II and Fig. 4, with average fluorescence lifetimes of about 1-2 ns, almost all excited molecules will go back to the ground state after 15-20 ns. Since we do not know the exact FRET rate between fluorophores in our experiments, we have to assume that the energy transfer can occur in any time between a few to about 20 ns, i.e. as long as we observe photons after a single laser pulse. Additionally, we should take into account the relaxation time of the acceptor molecule. When the acceptor is excited, it cannot receive another signal from the donor[3]. Thus, the acceptor should first pass the signal via another FRET to other molecules or emit a photon. It adds another 20 ns to the total delay of the signal transfer. Consequently, in this case the transmission rate should not exceed 1 bit per 40 ns, i.e. 25 Mbit/s. Moreover, as discussed earlier, the fluorophore excitation lifetimes may additionally vary because of the influence of neighboring molecules, what should be also taken into account when setting the transmission bit rate.

TABLE III

RESULTS OF THE FLIM MEASUREMENTS AND CALCULATIONS

| scenario | A | B | C | D |
|---|---|---|---|---|
| n (DoL of donor molecules) | 7 | 7 | 6-7 | 5-6 |
| m (DoL of acceptor molecules) | 6-7 | 5-6 | 5-6 | 5-6 |
| **measured $E_{(1,m)}$ [%]** | **25 ± 2** | **7 ± 2** | **36 ± 5** | **39 ± 6** |
| respective $BER_{(1,m)}$ | 0.375 | 0.465 | 0.32 | 0.305 |
| $R_0$ [nm] | 5.62 | 5.62 | 8.15 | 8.15 |
| av. donor-acceptor separation [nm] | 9 | 12 | 13 | 12 |
| **theoretical range for $E_{(1,m)}$ [%]** | **26-29** | **5-6** | **23-27** | **33-37** |

The probability that at least one of many donors transfers its excitation energy to at least one acceptor is much higher than having a single donor only, as explained in Section II. We can calculate FRET efficiency and BER for $(n,m)$ case, comparing equations (3) with (6) and (4) with (7):

$$E_{n,m} = 1 - \left(1 - E_{1,m}\right)^n , \qquad (11)$$

and:

$$BER_{n,m} = 0.5 \cdot \left(2 \cdot BER_{1,m}\right)^n . \qquad (12)$$

Consequently, on the basis of the measured values of $E_{(1,m)}$, we can predict the performance of the full $(n,m)$ MIMO-FRET channels. The results of those are given in Table IV. For scenarios C and D we have some ranges, as the number of donor dyes varies. It is worth noting that bit error rates at the level of 2-4% are easily achievable, even for a distance of 13 nm. The exception is the scenario B, where, however, the distance between the donor and acceptor sides is more than twice the length of the respective Förster distance.

In order to give the reader an even broader view as to what bit error rates can be achieved in MIMO-FRET channels, we also provide full BER curves for dye pairs being the subject of our experiments (549-594, 594-649), as well as for a few other DyLight dyes, well suited to nanocommunication. The curves are presented in Figure 6; the respective Förster distances and the numbers of donors/acceptors are also given. As we can see, using multiple acceptors increases the range where FRET-based communication can be used effectively. If we assume the required BER level to be 0.1%, the transmission range is about 35% longer in channels (1,6) compared to (1,1). Full MIMO communication is, however, even more efficient: in channels (6,6), the transmission range is 3.5 times longer than in channels (1,1). As a rule of thumb, it is worth remembering that for each donor-acceptor pair, the BER < 0.1% can be achieved at a distance equal to $R_0$, if a FRET channel (4,4) or larger is used (see Eq. (7)).

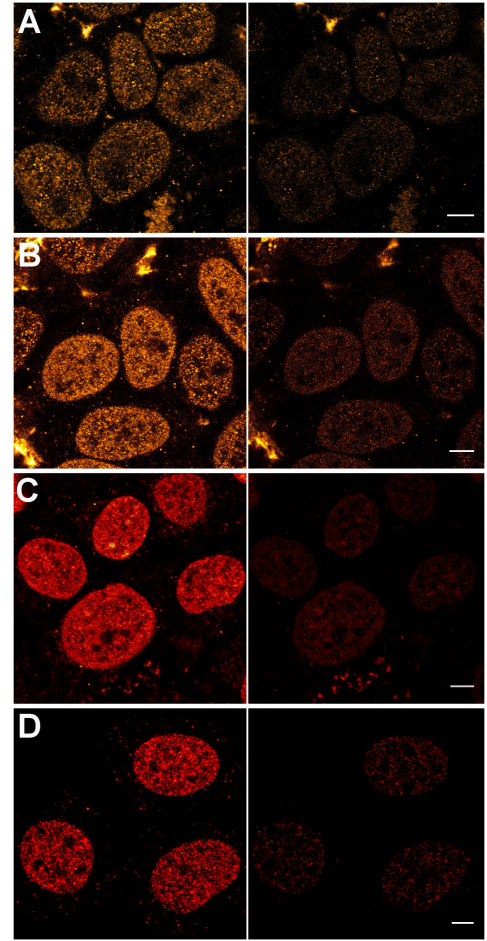

Fig. 5. The decrease of fluorescence intensity being the result of FRET in scenarios A, B, C and D. Two images for each scenario are shown: one before (left column) and one after (right column) staining with acceptor-bound secondary antibody. In all configurations, addition of the acceptor resulted in the decrease of fluorescence intensity of the donor. Scale bar: 5 um.

As FRET transmissions are so strictly limited in their range ($E$ decreases with the sixth power of the distance), the ability to create multi-hop connections is critical when thinking about future nanonetworks. Fluorophores have their emission spectra shifted a little towards the higher wavelengths compared to their respective absorption spectra [14], thus it is generally quite difficult to send a signal via FRET among identical fluorophores (there are some exceptions, see homo-FRET [14]). Instead, chains of spectrally matched molecules can be created. From Fig. 6, we see that a 6-element DyLight chain: 405→488→549→594→649→680 may effectively transfer a signal over a distance of several dozen nanometers (assuming each hop is equal to $R_0$, see the rule of thumb above). Such chains may be built not only with DyLight molecules, but also with GFP and its derivatives, Alexa Fluor, CyDye, LI-COR dyes, etc.

---

[3] In a scenario with multiple acceptors (MIMO-FRET), one might not wait for the excited acceptor to release its energy, but instead, to transfer the signal to other acceptor molecules. It, however, would complicate the decoding process and decrease the FRET efficiency (as the number of available acceptors is lower).

TABLE IV
FRET EFFICIENCIES AND BER VALUES FOR (1,M) AND (N,M) MIMO-FRET
CHANNELS

| Scenario | A | B | C | D |
|---|---|---|---|---|
| n (DoL of donors) | 7 | 7 | 6-7 | 5-6 |
| m (DoL of acceptors) | 6-7 | 5-6 | 5-6 | 5-6 |
| measured $E_{(1,m)}$ [%] | 25 ± 2 | 7 ± 2 | 36 ± 5 | 39 ± 6 |
| respective BER$_{(1,m)}$ | 0.375 | 0.465 | 0.32 | 0.305 |
| calculated $E_{(n,m)}$ [%] | 86.7 | 40 | 93-96 | 91-95 |
| calculated BER$_{(n,m)}$ | 0.067 | 0.3 | 0.02-0.04 | 0.025-0.04 |

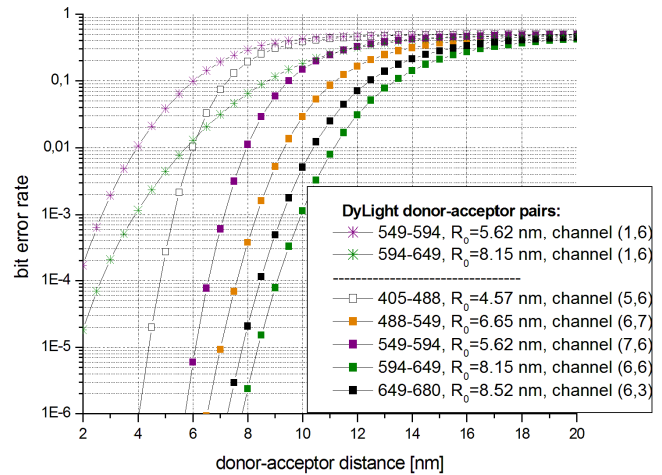

Fig. 6. Bit error rate curves for the chosen MIMO-FRET channels based on DyLight dyes.

## VI. CONCLUSIONS AND FUTURE WORKS

In this article, we have investigated the Förster Resonance Energy Transfer as a means of nanocommunication. The FRET phenomenon is characterized by a very short timescale, in the order of nanoseconds, in which a signal is non-radiatively transferred between molecules. Consequently, transmissions via FRET have much smaller propagation delays than other techniques considered for nanocommunications (even 4-5 orders of magnitude compared with calcium signaling, flagellated bacteria, molecular or catalytic motors) and a potential throughput of tens or even hundreds of Mbit/s.

The main drawbacks of FRET-based nanocommunication lie in its high bit error rate and short transmission ranges. While the former issue can be solved using multiple donors and acceptors, the latter requires a careful design of FRET communication channels, especially by choosing for transmitter and receiver antennas the pairs of fluorescent dyes characterized by large Förster distances and high degrees of labeling. In this paper, we have proposed to use a fluorescent family of DyLight dyes which fits very well the abovementioned criteria. We have constructed 4 different nanonetworks based on Immunoglobulin G antibodies and DyLight dyes and experimentally validated the communication efficiency between their nodes. The experiments confirm successful communication over the distances of 12-13 nm, which is a 50% increase comparing with the results previously reported in the literature. Higher ranges can be reached via multi-hop connections making use of fluorophore chains.

While these research results are very promising, many questions still remain. Importantly, we do not know how future nanomachines will communicate with their nanoantennas (the fluorophore molecules) and send signals to them. One option is Bioluminescence Resonance Energy Transfer (BRET), where a nanomachine may induce a chemical reaction generating energy that excites donors. In order to fully exploit MIMO-FRET communication, all donors should be excited simultaneously, which, however, may not be straightforward to achieve.

The experimentally measured FRET efficiencies are in reasonably good agreement with theoretical values, but more exact assessments could be made if the distances between the fluorophores were controlled with higher precision. These could be obtained with the aid of molecular biology and genetic engineering techniques using polymers [26]. Polymers are chains of biological molecules composed of many repeated sub-units; the length of one polymer can easily reach 1000 nm, whereas a single element can be 1 nm or less. In living organisms we can find a few types of polymers: proteins (e.g. actin filaments, microtubules), polysaccharides (e.g. starch) and nucleic acids (DNA and RNA). If sub-units of a polymer were tagged with fluorescent molecules, the distances between these fluorophores could be determined with high precision: about 1 nm or better. Additionally, a polymer molecule can be composed of a main chain with one or more side chains or branches, which may be an interesting scenario considering possible research on signal routing for FRET-based nanocommunication.

## REFERENCES

[1] B. Fadeel, Nanosafety: towards safer design of nanomachines, Journal of Internal Medicine, 2013, 274, 578-580.

[2] J. Gupta, Nanotechnology applications in medicine and dentistry, Journal of Investigative and Clinical Dentistry, 2011, 2, 81-88.

[3] K. Kim, J. Guo, X. Xu, D. Fan, Recent Progress on Man-Made Inorganic Nanomachines, Small, 2015, DOI: 10.1002/smll.201500407.

[4] T. Nakano, T. Suda, M. Moore, R. Egashira, A. Enomoto, K. Arima, Molecular communication for nanomachines using intercellular calcium signaling, Proceedings of the Fifth IEEE Conference on Nanotechnology, June 2005, pp. 478–481.

[5] T. Nakano, A.W. Eckford, T. Haraguchi, Molecular Communications, Cambridge University Press, 2013.

[6] M. Gregori, I.F. Akyildiz, A New NanoNetwork Architecture using Flagellated Bacteria and Catalytic Nanomotors, IEEE Journal of Selected Areas in Communications, vol. 28, pp. 612-619, May 2010.

[7] M. Moore, A. Enomoto, T. Nakano, R. Egashira, T. Suda, A. Kayasuga, H. Kojima, H. Sakakibara, K. Oiwa, A design of a molecular communication system for nanomachines using molecular motors, Proceedings of the Fourth Annual IEEE International Conference on Pervasive Computing and Communications, March 2006.

[8] V. Serreli, C. Lee, E.R. Kay, D.A. Leigh, A molecular information ratchet, Nature 445: 523–527, 2007.

[9] L. Parcerisa, I.F. Akyildiz, Molecular Communication Options for Long Range Nanonetworks, Computer Networks (Elsevier) Journal, vol. 53, pp. 2753-2766, 2009.

[10] V. Loscri, A.M. Vegni, An Acoustic Communication Technique of Nanorobot Swarms for Nanomedicine Applications, IEEE Transactions on NanoBioscience, vol. 14, no. 6, pp. 598-607, September 2015.

[11] M. Kuscu, O.B. Akan, A Physical Channel Model and Analysis for Nanoscale Communications with Förster Resonance Energy Transfer (FRET), IEEE Transactions on Nanotechnology, vol. 11, no. 1, pp. 200-207, January 2012.

[12] M. Kuscu, O.B. Akan, Multi-Step FRET-Based Long-Range Nanoscale Communication Channel, IEEE Journal on Selected Areas in Communications, vol. 31, no. 12, pp. 715-725, December 2013.

[13] K. Wojcik, K. Solarczyk, P. Kulakowski, Measurements on MIMO-FRET nanonetworks based on Alexa Fluor dyes, IEEE Transactions on Nanotechnology, vol. 14, no. 3, pp. 531-539, May 2015.

[14] Joseph R. Lakowicz, Principles of Fluorescence Spectroscopy 3 ed., Springer, 2006.

[15] Fábián ÁI, Rente T, Szöllosi J, Mátyus L, Jenei A., Strength in numbers: effects of acceptor abundance on FRET efficiency. Chemphyschem. 2010 Dec 3;11(17):3713-21.

[16] I.E. Telatar, Capacity of Multi-antenna Gaussian Channels, AT&T Bell Laboratories, Technical Memorandum, June 1995.

[17] G.J. Foschini and M.J. Gans, On limits of wireless communications in a fading environment when using multiple antennas, Wireless Personal Communications, vol. 6, pp. 311–335, March 1998.

[18] DyLight Fluors – Technology and Product Guide, http://www.piercenet.com/guide/dylight-fluors-technology-product-guide

[19] P. Sarkar, S. Sridharan, R. Luchowski, S. Desai, B. Dworecki, M. Nlend, Z. Gryczynski, I. Gryczynski, Photophysical properties of a new DyLight 594 dye, Journal of Photochemistry and Photobiology B: Biology, Volume 98, Issue 1, 21 January 2010, Pages 35-39.

[20] S. D'Auria, E. Apicella, M. Staiano, S. Di Giovanni, G. Ruggiero, M. Rossi, P. Sarkar, R. Luchowski, I. Gryczynski, Z. Gryczynski, Engineering resonance energy transfer for advanced immunoassays: The case of celiac disease, Analytical Biochemistry, Volume 425, Issue 1, 1 June 2012, Pages 13-17.

[21] Yih Horng Tan, Maozi Liu, Birte Nolting, Joan G. Go, Jacquelyn Gervay-Hague, Gang-yu Liu, A Nanoengineering Approach for Investigation and Regulation of Protein Immobilization, ACS Nano 2 (11), pp. 2374–2384, 2008.

[22] Jeremy M. Berg, John L. Tymoczko, Lubert Stryer, Biochemistry, 5th Edition. New York, W.H. Freeman and Company, 2002.

[23] Mikhail Y. Berezin, Samuel Achilefu, Fluorescence Lifetime Measurements and Biological Imaging, Chem. Rev., 2010, 110(5), pp. 2641–2684.

[24] William E. Paul, Fundamental Immunology 7th Ed., Wolters Kluwer Health/Lippincott Williams & Wilkins, 2013.

[25] M. Kuscu, O.B. Akan, FRET-Based Nanoscale Point-to-Point and Broadcast Communications with Multi-Exciton Transmission and Channel Routing, IEEE Transactions on NanoBioscience, vol. 13, no. 3, pp. 315-326, September 2014.

[26] M. Rubinstein, R.H. Colby, Polymer physics. Oxford, New York, Oxford University Press, 2003.

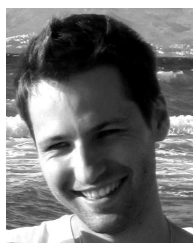

**Kamil Solarczyk** received the M.Sc. degree in biophysics from the Jagiellonian University, Kraków, Poland, in 2010. Since then, he has been working toward the Ph.D. degree in the Faculty of Biochemistry, Biophysics and Biotechnology. His research interests include the DNA repair processes, chromatin architecture and molecular communication.

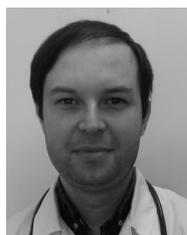

**Krzysztof Wojcik** received the M.Sc. and Ph.D. degrees in biophysics from the Jagiellonian University in Krakow, Poland 2003 and 2015, respectively; and M.D. from the Jagiellonian University Medical College in Kraków, Poland 2007. He was an Assistant at Division of Cell Biophysics Faculty of Biochemistry, Biophysics and Biotechnology Jagiellonian University (2007-2014). He is an Assistant at Allergy and Immunology Clinic in II Chair of Internal Medicine CMUJ. His research interests include the fields of confocal microscopy techniques and its application in autoantibodies research and use of fluorescent labeled antibodies in nanocommunications.

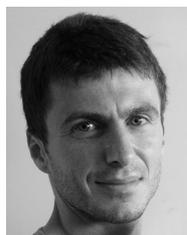

**Pawel Kulakowski** received the Ph.D. degree in telecommunications from the AGH University of Science and Technology in Krakow, Poland, in 2007. Currently he is working there as an Assistant Professor. He was also working few years in Spain, as a visiting post-doc or professor at Technical University of Cartagena, University of Girona, University of Castilla-La Mancha and University of Seville. He co-authored about 30 scientific papers, in journals, conferences and as technical reports. He was also involved in numerous research projects, especially European COST Actions: COST2100, IC1004 and CA15104 IRACON, focusing on topics of wireless sensor networks, indoor localization and wireless communications in general. His current research interests include molecular communications and nano-networks. Dr. Kulakowski was recognized with several scientific distinctions, including 3 awards for his conference papers and a scholarship for young outstanding researchers.